\documentclass[aps,pra,twocolumn,superscriptaddress,showpacs]{revtex4}
\usepackage[T1]{fontenc}
\usepackage{amsmath}
\usepackage{bm}
\usepackage{natbib}
\usepackage{graphicx} 
\usepackage{epsfig}

\def \dV {\mathrm{d}^3 r}
\def \E {\mathcal{E}}

\begin{document}


\title{Scattering 
of dark-state polaritons in optical lattices and quantum phase gate for photons}
\author{M. Ma\v{s}alas}
\affiliation{Technische Universit\"at Kaiserslautern, 67663 Kaiserslautern, 
Germany}
\affiliation{Vilnius University Research Institute of Theoretical Physics and Astronomy, 2600 Vilnius, Lithuania}
\author{M. Fleischhauer}
\affiliation{Technische Universit\"at Kaiserslautern, 67663 Kaiserslautern, 
Germany}
\date{\today}


\begin{abstract}
We discuss the quasi 1-D scattering of two counter-propagating, 
dark-state
polaritons (DSP), each containing a single excitation. DSPs are formed from
photons in media with electromagnetically induced transparency and 
are associated with ultra-slow group velocities.   
State-dependent elastic collisions
of atoms at the same lattice site lead to a nonlinear interaction.
It is shown that the scattering process in a deep optical 
lattice filled by cold atoms generates a large and homogeneous conditional 
phase shift between two individual polaritons. 
The latter has potential applications
for a photonic phase gate. The quasi 1-D scattering problem 
is solved analytically and the 
influence of degrading processes such as dephasing due to 
collisions with ground-state atoms is discussed.
\end{abstract}
\maketitle




A major challenge for quantum information processing using individual
photons as qubits is the implementation of 
logic operations.
Such operations require efficient nonlinear interactions 
for pairs of photons
which cannot be achieved in conventional optical materials. 
As the effect of the photon-photon 
coupling depends on the nonlinear susceptibilities as well as the
interaction time, it has been suggested to use ultra-slow light in resonant
systems with electromagnetically induced transparency (EIT), where  
the interaction time is long and the nonlinear susceptibilities become 
large due to resonance enhancement \cite{EIT,resNLO,Harris}. 
At the boundary of an stationary 
EIT medium light pulses become spatially compressed in the propagation
direction by the ratio of group velocity $v_{\rm gr}$  to the
vacuum speed of light \cite{Harris}. As a consequence the number of photons in the pulse
decreases by the same factor. Excitations
are  temporarily transferred to the
medium by the formation of quasi-particles, so-called dark-state
polaritons (DSP) which are superpositions
of electromagnetic and  atomic degrees of freedom \cite{FleischhauerPRL-2000}. 
As another consequence of the pulse compression 
the interaction time in a head-on collision
stays constant irrespective of the value of $v_{\rm gr}$. Thus in order to
achieve long interaction times co-propagating pulses were considered 
\cite{LukinPRL:00}.
In this case the interaction is however not 
homogeneous and it is difficult to avoid spectral broadening of the
wave packet.

We here suggest a completely different mechanism for an efficient
nonlinear interaction between ultra-slow light pulses. The
slow-down corresponds to a shift of the
polariton composition from pure photons
to matter-waves \cite{FleischhauerPRL-2000}. 
Furthermore the pulse compression leads to an increasing
density of the matter component. 
Thus collisional interactions 
between atomic excitations can yield an effective nonlinear
coupling between two wave-packets. To further enhance the strength of this
interaction we consider a lattice potential in the tight-binding limit. 
Starting from a fully quantized effective one-dimensional
model of light propagation in a lattice
we derive analytic solutions for the quantum scattering of two
single-photon wave-packets in the s-wave scattering limit. We show that
the pulses attain a homogeneous conditional phase shift which may be
large enough for the implementation of a quantum phase gate.




\begin{figure}[ht]
\includegraphics[width=6.5cm]{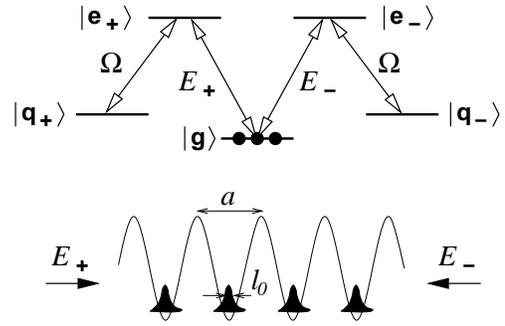} 
\caption{{\it top:} Atomic five-level system with quantized probe fields $E_\pm$
of opposite circular polarization and propagation direction. $\Omega$
denotes the Rabi-frequency of the classical, undepleted  
control field.
{\it bottom:} The atoms are assumed to be confined in a 3-D lattice potential
with lattice constant $a$. In the tight-binding regime they occupy only the 
lowest Wannier state of effective width $l_0$.}
\label{atomic-system}
\end{figure}

Let us consider a cold gas of bosonic, five-level atoms as shown in 
Fig.~\ref{atomic-system} in a deep three-dimensional lattice potential 
under tight-binding conditions. The atoms form an $M$-type system, 
with the ground state $|g\rangle$ and the excited states $|e_\pm\rangle$ 
coupled by two orthogonal polarizations of a quantized probe field $\hat E_\pm$
propagating in the $+z$ or $-z$ direction respectively. 
The excited states are furthermore coupled
to meta-stable state $|q_\pm\rangle$ by a classical probe field
of Rabi-frequency $\Omega$. All atoms are initially in the ground
state $|g\rangle$. The atoms are described by five  Bose
fields $\psi_i$, where $i\in\{g,e_\pm,q_\pm\}$ denotes the internal state. 
The Hamiltonian of the system is given by
$ H=H_{\rm at}+H_{\rm lat}+H_{\rm at-f}+H_{\rm coll}$,
where the atomic part reads
\begin{equation}
H_{\rm at} = 
\sum_{i=g,e_\pm,q_\pm} { \int{\dV \psi_{i}^{\dagger} \left(-\frac{\hbar^2}{2m}
\nabla^2 +\hbar\omega_{i}\right)\psi_{i}} }.
\label{eq:ham-at}
\end{equation}
$H_{\rm lat}$ is the lattice potential, which is assumed to be the
same for all internal states.
The interaction with e.m. field reads in rotating-wave-approximation
\begin{eqnarray}
H_{\rm at-f} &=& -\int{\dV
\sum_{i=+,-}
\psi_{e_i}^{\dagger}
\left({\mu}E_i^{(+)}\left(\mathbf{r}\right)\right)\psi_{g}} \nonumber\\
&&- \int{\dV 
\sum_{i=+,-}
\psi_{e_i}^{\dagger}\hbar\Omega 
\left(\mathbf{r}\right)\psi_{q_i}} + \rm{H.c}.
\label{eq:ham-atfield}
\end{eqnarray}
Finally we take into account collisions between atoms in 
the internal states $|g\rangle$ and $|q_\pm\rangle$ 
as well as collisions between atoms in $|q_-\rangle$ and $|q_+\rangle$
in s-wave approximation
\begin{eqnarray}
H_{\rm coll} 
&=& \frac{\hbar u_{g}}{2}
\int \dV\,  \psi_{g}^{\dagger}\psi_{g}^{\dagger} \psi_{g}\psi_{g}
\nonumber\\
&&+\sum_{i,k=+,-} \frac{\hbar u_{ik}}{2}
\int \dV \,\psi_{q_i}^{\dagger}\psi_{q_k}^{\dagger} 
\psi_{q_k}\psi_{q_i}
\nonumber\\
&& +\sum_{i=+,-}
\hbar u_{gi}\int \dV\, \psi_{q_i}^{\dagger} \psi_{g}^\dagger\psi_{g}\psi_{q_i}.
\label{eq:ham-coll}
\end{eqnarray}
Collisions of atoms in the excited state are irrelevant as these states
will attain vanishing small population. 
Here $E^{(+)}_\pm\left(\mathbf{r}\right)$ are the positive frequency parts 
of the probe field operators corresponding to the two
orthogonal polarized modes, $\Omega\left(
\mathbf{r}\right)$ is the Rabi frequency of the control field,
and the  $u$'s describe the
collision strength, which can be expressed in terms of the 
corresponding $s$-wave scattering length $u_i=4\pi a_i \hbar/m$.
 The $\omega_{i}$'s are the frequencies corresponding to the 
electronic energy levels.

Using eqs.\eqref{eq:ham-at}-\eqref{eq:ham-atfield}, and
\eqref{eq:ham-coll} the Heisenberg equations for the field operators can 
easily be obtained:
\begin{subequations}
\label{eq:psi}
\begin{align}
&i\hbar\frac{\partial \psi_g}{\partial t} = \left(-\frac{\hbar^2}{2m}
\nabla^2 +V+\hbar\omega_{g} \right)\psi_{g} -\mu^* E_-^{(-)}
\psi_{e_-}
\nonumber\\
&\quad -\mu^* E_+^{(-)}
\psi_{e_+} + \hbar\Bigl(u_{g} \psi_g^\dagger\psi_g
+ \sum_{i=+,-} u_{gi} \psi_{q_i}^\dagger\psi_{q_i}\Bigr)\psi_g, 
\label{eq:psi_g} \\
%
%
&i\hbar\frac{\partial \psi_{e_\pm}}{\partial t} = \left(-\frac{\hbar^2}{2m}
\nabla^2 +V+\hbar\omega_{e\pm} \right)\psi_{e_\pm} -\mu E_\pm^{(+)}\psi_{g}
\nonumber\\
&\qquad-\hbar\Omega\psi_{q_\pm},
\label{eq:psi_e} \\
%
%
&i\hbar\frac{\partial \psi_{q_\pm}}{\partial t} = \left(-\frac{\hbar^2}{2m}
\nabla^2 +V+\hbar\omega_{q\pm} \right)\psi_{q_\pm} -\hbar\Omega^{\ast} 
\psi_{e_\pm}\nonumber\\ &\quad 
+ \sum_{k=+,-}\hbar\Bigl(u_{gk}\psi^{\dagger}_g\psi_{g}
+ u_{{\pm}k}\psi_{q_k}^\dagger\psi_{q_k}\Bigr)\psi_{q_\pm}.
\label{eq:psi_q}
\end{align}
\end{subequations}
We here have not included decay from the excited states as these states
will attain only negligible population. 
The above operator
equations are nonlinear and thus 
impossible to solve exactly. Therefore 
approximations are needed.

First it is  assumed that both probe and control fields propagate along
the $z$ axis and can be written in the following form:
$E_\pm^{(+)}\left(\mathbf{r},t\right) = \E_\pm
\left(\mathbf{r},t\right) e^{i\left(\pm kz-\omega t\right)}, 
\Omega\left(\mathbf{r},t\right) = \Omega_{0}\left(t\right)
e^{-i\omega_{c} t}$.
Here both $\E_\pm \left(\mathbf r,t\right)$ and $\Omega_{0}
\left(t\right)$ are slowly varying functions of $\mathbf r$ and $t$.
One also assumes that the probe field coupling  $\mu\E_\pm/\hbar$ 
is much weaker than the control field coupling $\Omega_0$. In this limit we may
consider the control field undepleted and classical and can thus set
$\Omega_0$ constant. 
The assumption that all atoms were initially in the ground
state $g$ then also implies
that only a small fraction of atoms is excited
to the states $q_\pm$.  

The strength of the lattice is considered to be large 
enough such that 
only the lowest energy level of each potential well is occupied and
tunneling between the wells is negligible.  
In this limit the atomic field operators can be expanded in the basis of
(real) Wannier functions \cite{JakschPRL-1999}. 
Since only the lowest lattice state is occupied, only
Wannier states of the lowest Bloch band 
$W_j(\mathbf{r})=W_0(\mathbf{r}-\mathbf{r}_j)$ survive in this 
expansion. In a deep lattice Wannier-functions of neighboring sites
have only negligible overlap and one has
$\int\dV W_i(\mathbf{r}) W_k(\mathbf{r}) =\delta_{ik}$.
This expansion along with a separation of fast oscillating terms yields
$\psi_{g} = \sum_{j}{ W_{j}\left(\mathbf{r}\right)g_{j}\left(t\right)}$ ,
$\psi_{e_\pm} = \sum_{j}{ W_{j}\left(\mathbf{r}\right) e_{\pm j}\left(t\right)
e^{i\left(\pm kz-\omega t\right)} }$, and $
\psi_{q_\pm} = \sum_{j}{ W_{j}\left(\mathbf{r}\right) q_{\pm j}\left(t\right)
e^{\pm i\left(k-k_{c}\right)z -i\left(\omega -\omega_{c}\right) t} }$ .
Here the operators $g_{j}$, $e_{\pm j}$, and $q_{\pm j}$ are slowly varying 
in $t$ and the summation runs over all lattice sites.

Substituting the Wannier-expansion into
eqs.\eqref{eq:psi} yields equations for the slowly varying atomic
variables. Since the Wannier functions are well localized, the slowly varying
functions remain almost constant within one lattice site. 
\begin{subequations}
\label{eq:motion}
\begin{eqnarray}
&&\dot g_{j} = i\frac{\mu}{\hbar}\E_-^\dagger\left(\mathbf{r}_j\right) e_{-j}
+i\frac{\mu}{\hbar}\E_+^\dagger\left(\mathbf{r}_j\right) e_{+j}\nonumber\\
&&\quad-i\Bigl(u_g g_j^\dagger g_j +u_{g+} q_{+j}^\dagger q_{+j}
+u_{g-} q_{-j}^\dagger q_{-j}\Bigr)\frac{f^3}{a^3} g_j\qquad
\label{eq:G_j}
\end{eqnarray}
\begin{equation}
\dot e_{\pm j} = -i\Delta_{e\pm} e_{\pm j}
+i\frac{\mu}{\hbar}\E_\pm \left(\mathbf{r}_j\right)g_{j} +i\Omega_{0}q_{\pm j}
\label{eq:E_j}
\end{equation}
\begin{eqnarray}
&&\dot q_{\pm j} = -i\Delta_{q\pm} q_{\pm j}
+i\Omega_{0}^{\ast}e_{\pm j} \quad\label{eq:Q_j}\\
&&\quad -i\Bigl(u_{\pm\pm} q_{\pm j}^{\dagger}q_{\pm j}
+u_{+-} q_{\mp j}^{\dagger}q_{\mp j}
+u_{g\pm} g_j^\dagger g_j\Bigr) \frac{f^3}{a^3} q_{\pm j}\qquad\nonumber
\end{eqnarray}
\end{subequations}
with the detunings
%
%
$\Delta_{e\pm}=\omega_{e\pm}-\omega +{\hbar k^2}/{2m}$ ,
%
%
$\Delta_{q\pm}=\omega_{q\pm}-\omega
+\omega_{c} +{\hbar \left(k-k_{c}\right)^2}/{2m}.$
%
%
We here have set $\omega_g=0$. The factor $f=a/l_0$ describes the
confinement strength in the lattice and is defined as
$\int \dV\,  W_i(\mathbf{r})^4 =f^3/a^3$.

In order to solve the above equations of motion for the matter-field
operators a 
weak-probe approximation will be applied. 
In zeroth order of the probe field the states $|e_\pm\rangle$ and
$|q_\pm\rangle$ remain unpopulated and one finds for the ground state operators
\begin{equation}
\dot g_{j} = -i u_g \frac{f^3}{a^3} g_j^\dagger g_j g_j
\label{eq:G_j-2}
\end{equation}
If the lattice has a regular filling with a well defined 
number $N_j=N$ of atoms per site, we can make the replacement
$g_j^\dagger g_j \to N$ in eq.(\ref{eq:G_j-2}). A regular filling
can be achieved e.g. by employing a Mott-insulator transition
in a lattice \cite{JakschPRL-1999,Bloch-exp}. 
In this case the self-phase modulation described by
\eqref{eq:G_j-2} can just be absorbed in the definition of the
energy of the ground state atoms. Furthermore we can 
replace the ground-state operator in the equations for the
operators of the other states by a c-number $g_j\to \sqrt{N}$.
With this we obtain in first order of the probe field
\begin{subequations}
\begin{equation}
\dot e_{\pm j} = -i\Delta_{e\pm} e_{\pm j}
+i\frac{\mu\sqrt{N}}{\hbar}\E_\pm\left(\mathbf{r}_j\right) 
+i\Omega_{0}q_{\pm j}
\label{eq:E_j-1}
\end{equation}
\begin{eqnarray}
&&\dot q_{\pm j} = -i(\Delta_{q\pm }+ u_{g\pm }f^3 n) q_{\pm j}
+i\Omega_{0}^{\ast}e_{\pm j} \nonumber\\
&&\quad -i u_{\pm\pm} \frac{f^3}{a^3} q_{\pm j}^{\dagger}q_{\pm j} q_{\pm j}
 -i u_{+-} \frac{f^3}{a^3} q_{\mp j}^{\dagger}q_{\mp j} q_{\pm j}\qquad
\label{eq:Q_j-1}
\end{eqnarray}
\end{subequations}
where $n=N/a^3$ is the average density of atoms .
Next we assume resonance conditions, i.e.
$\Delta_{e\pm}=0$ and $\Delta_{q\pm}+u_{g\pm} n f^3 =0$ and 
that the probe field varies sufficiently slowly. Under these conditions
we can apply an adiabatic approximation. 

In zeroth order of the adiabatic approximation the time derivative 
in the equations of motion for the atomic variables
\eqref{eq:motion} is neglected which yields
\begin{subequations}
\label{eq:zero-sol}
\begin{align}
q_{\pm j}\left(t\right) &= -\frac{\mu\sqrt{N}}{\hbar\Omega_{0}} 
\E_{\pm j}
\label{eq:Q-zero} \\
%
e_{\pm j}\left(t\right) &= \frac{u_{\pm\pm} f^3}{\Omega_{0}^{\ast}a^3}
q_{\pm j}^{\dag}q_{\pm j}q_{\pm j} 
+\frac{u_{+-} f^3}{\Omega_{0}^{\ast}a^3}
q_{\mp j}^{\dag}q_{\mp j}q_{\pm j} 
\label{eq:E-zero}
\end{align}
\end{subequations}
with $\E_{\pm j} \equiv \E_\pm\left(\mathbf{r}_j\right)$. 
Thus 
\begin{eqnarray}
e_{\pm j}^{(0)}\left(t\right) &=& 
-\hbar u_{\pm \pm} f^3\frac{\mu\sqrt{N}}
{\left|\hbar\Omega_{0}\right|^2} \frac{|\mu|^{2}n}
{\left|\hbar\Omega_{0}\right|^2} \E_{\pm j}^{\dag}\E_{\pm j}\E_{\pm j}
\nonumber\\
&&-\hbar u_{+-} f^3\frac{\mu\sqrt{N}}
{\left|\hbar\Omega_{0}\right|^2} \frac{|\mu|^{2}n}
{\left|\hbar\Omega_{0}\right|^2} \E_{\mp j}^{\dag}\E_{\mp j}\E_{\pm j}
\label{eq:E-sol}
\end{eqnarray}
Proceeding in the same manner with the first order of the 
adiabatic approximation we
find that the operators corresponding to the excited states $|e_\pm\rangle$ 
contain a term proportional to the time derivative of the 
probe fields 
\begin{equation}
e_{\pm j}^{(1)}\left(t\right) = e_{\pm j}^{(0)}\left(t\right)
+\frac{i\mu \sqrt{N}}{\hbar|\Omega_{0}|^{2}} 
\frac{\partial}{\partial t} \E_{\pm j} 
\label{eq:E-sol-first}
\end{equation}
Here higher order terms in $\E_{\pm j}$ containing a time derivative
were neglected as they correspond to corrections higher order in both the
probe field and the adiabaticity parameter.

The adiabatic solutions for the matter fields can now be used to
calculate the slowly varying amplitude of the probe-field polarizations
$P_{\pm}(\mathbf{r},t) =
\mu^{\ast}\sum_{j}{|W_{j}(\mathbf{r})|^2 g_{\pm j}^{\dagger}(t)e_{\pm j}(t)}.
$
%
%
%
%
Since the Wannier functions are strongly localized around the center of the
potential wells, the microscopic polarization 
changes rapidly in space. On the another hand in Maxwell's equations
for the electric field only the macroscopic polarization
enters. The macroscopic polarization can be obtained 
by averaging over a volume small compared to the
wavelength. If the lattice constant $a$ is sufficiently smaller than
the relevant wavelength of the probe field, the lattice structure
disappears in the polarization. In the opposite limit effects like 
Bragg scattering can occur, but we are not interested in these phenomena
here and thus assume $a<\lambda$. 

Using the macroscopic polarization we find the following equation of motion
for the slowly varying amplitude of the field operator
\begin{eqnarray}
\left(\frac{\partial}{\partial t} \pm  v_{gr}\frac{\partial}{\partial z}\right)
\E_\pm &=&
 -iu_{\pm\pm}\frac{\lambda \varepsilon_0}{\hbar \pi v_{\rm gr}}f^{3}
\E_\pm^{\dag}\E_\pm\, \E_\pm\nonumber\\
&& -iu_{+-}\frac{\lambda \varepsilon_0}{\hbar \pi v_{\rm gr}}f^{3}
\E_\mp^{\dag}\E_\mp\, \E_\pm
\label{eq:field-final}
\end{eqnarray}
with group velocity $v_{gr} =
2c\hbar|\Omega_{0}|^{2}\varepsilon_{0}/|\mu|^{2}\omega n$ 
(assuming $v_{\rm gr}\ll c$). 
 The appearance of the group velocity in the
denominator of \eqref{eq:field-final} suggest on first glance an 
diverging nonlinear interaction when the group velocity approaches zero.
One should take into account however that due the pulse compression the
total photon number $N_{\rm ph}\sim \int \dV\, \E^\dagger\E$ is only
a fraction $v_{\rm gr}/c$ of the input value. \eqref{eq:field-final} becomes
much more transparent if it is translated into an equation of motion
of the dark-state polaritons \cite{FleischhauerPRL-2000}
\begin{equation}
\Psi_\pm(z,t) = \cos\theta \tilde\E_\pm(z,t) 
-\sin\theta \sqrt{A} \tilde\psi_{q\pm}(z,t)
\end{equation}
where $\E =\tilde \E \sqrt{\hbar\omega/2\varepsilon_0 A}$, $A$ being the
cross section of the light beam, $v_{\rm gr} =c \cos^2\theta$, and
$\tilde\psi_{q\pm}$ are the slowly-varying amplitudes of the matter fields.. 
In the adiabatic limit considered here
the orthogonal quasi-particle, the bright-state polariton 
is not excited and thus $\tilde\E_\pm = \Psi_\pm \cos\theta$. 
With this we find
the following propagation equation inside the medium
%
%
\begin{eqnarray}
\left(\frac{\partial}{\partial t}\pm v_{gr}\frac{\partial}{\partial z}\right)
\Psi_\pm &=& 
-i\frac{2 a_{\pm\pm}\lambda v_{\rm rec}}{A}f^{3}
\Psi_\pm^{\dag}\Psi_\pm\Psi_\pm\qquad\nonumber\\
&& -i\frac{2 a_{+-}\lambda v_{\rm rec}}{A}f^{3}
\Psi_\mp^{\dag}\Psi_\mp\Psi_\pm,\label{eq:polariton-final}
\end{eqnarray}
where we have substituted the s-wave scattering length $a_{ij}$ and the
recoil velocity $v_{\rm rec}= \hbar\omega/mc$.
The polariton number densities $n_{\pm}=\Psi_\pm^\dagger(z)\Psi_\pm(z)$ 
undergo a sudden increase at the boundary of the medium since the electric
field is continuous there. Inside the medium they
propagate form-stable with
$v_{\rm gr}$.

\begin{figure}[ht]
\includegraphics[width=7.5 cm]{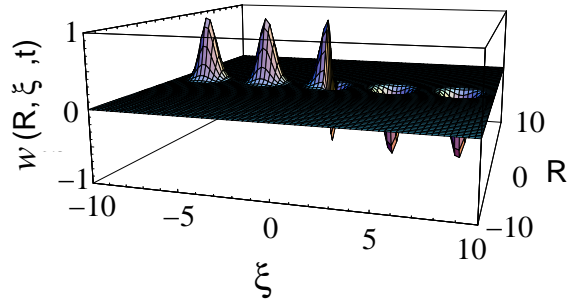} 
\caption{Snapshots of two-particle wave-function $w(R,\xi,t)$ at equidistant
times as function of center-of-mass coordinate $R$ and difference
coordinate $\xi=z-z^\prime$. $\xi<0$ ($\xi>0$) corresponds to 
polaritons propagating toward (away from) each other. Overlapping
components generate a phase flip $\Delta\phi$, which is here taken to be $\pi$.
}
\label{wavefunction-3D}
\end{figure}

Eq.\eqref{eq:polariton-final} leads to a self- and cross-phase modulation of
the dark-state polaritons. If initially only one polariton is excited
of each sort the self-phase modulation vanishes. To solve the quasi-1D
scattering problem it is convenient to introduce the
two-particle wave-function
\begin{equation}
w(z,z^\prime,t)= \langle 0| \Psi_+(z,t) \Psi_-(z^\prime,t) |\phi\rangle
\end{equation}
where $|\phi\rangle$ is the initial state vector of the system and $|0\rangle$
corresponds to the polariton vacuum. One can show that in the case considered
here, namely only one polariton of each class is initially excited, all
information is contained in $w$.  
In terms of center-of-mass and difference coordinates
$R=(z+z^\prime)/2$ and $\xi=z-z^\prime$ the equation of motion for $w$ reads:
\begin{equation}
\left(\frac{\partial}{\partial t} + 2 v_{gr} \frac{\partial}{\partial
\xi}\right) w
= -2 i\frac{\delta\left(\xi\right) a_{+-}\lambda v_{\rm rec} }{A} f^3
w.
\label{eq:two-particle}
\end{equation}
This equation has a simple interpretation. The l.h.s. describes the
propagation of the two components in opposite directions. The r.h.s.
describes an interaction for $z=z^\prime$, i.e. when the
two polaritons meet.
The interaction conserves the center-of-mass of the two polaritons
and the solution of \eqref{eq:two-particle} reads
\begin{equation}
w(R,\xi,t)=w\left(R,\xi-2 v_{\rm gr}t,0\right)
\exp\Bigl(-i\Delta\phi \, \Theta(\xi)\Bigr)
\label{eq:phase}
\end{equation}
where $\Theta(\xi)$ is the Heaviside step function.
One recognizes that the shape of the two photon wave-function remains unchanged
by the collisions and there is only a homogeneous collision phase. This is
illustrated for the case $\Delta\phi=\pi$ in Fig.\ref{wavefunction-3D}. 

The conditional phase shift between the polaritons which originate from two
single-photon wave-packets is then transferred back to two photons
at the exit of the medium. In this way a quantum phase gate between
two individual photons could be implemented, if the conditional phase
shift can reach the value of $\pi$.
The collision phase in \eqref{eq:phase} is given by
\begin{equation}
\Delta\phi= \frac{a_{+-} \lambda}{A} \frac{v_{\rm rec}}{v_{\rm gr}} f^3.
\end{equation}
It is interesting to note that $\Delta\phi$ does not depend on the pulse 
parameters. It is not necessary that the two pulses
have a certain length or the same shape. Both should however occupy the
same transverse mode. One also recognizes that the use of a lattice potential
has two important effects. First of all phase diffusion of the individual
dark-state polaritons caused by the scattering of atoms in states 
$|q_\pm\rangle$ with ground-state atoms is eliminated by the regular
filling. Secondly the local enhancement of the density leads to 
an enhancement factor $f^3$. In a deep lattice $f$ can be as large as 10.
To give an estimate of achievable phase shifts let us assume
$a_{+-}=10$nm, $v_{\rm gr}=10 v_{\rm rec}$, $A=\lambda^2$, 
$\lambda=800$nm, and $f=10$. This yields a phase shift on the order 
of unity, which is of the required order of magnitude.

The main limitation of the present scheme is set by the finite dephasing time
of the dark-state polaritons. For very small group velocities the interaction
time $T=L/2v_{\rm gr}$ of the pulses becomes rather long. Since the compressed
pulse length in the medium $L$ should be sufficiently larger than the 
wavelength, the minimum $T$ is about $10\lambda/v_{\rm gr}$ which for the
above parameters with $v_{\rm rec}\approx 1$ cm/sec gives
$T\approx 40\mu$sec. In slow-light and light-''stopping'' experiments
\cite{HauNature:99,HauNature:01,PhillipsPRL:01}
dephasing times of msec have been observed and with the use of a deep 
lattice even longer times should be achievable. 

In summary we have shown that scattering of 
ultra-cold atoms in a deep 3-D lattice
together with the transfer of excitations between
photons and atomic excitations through dark-state polaritons 
can be used for a conditional homogeneous phase shift between individual
photons. An essential requirement to obtain sufficiently large 
phase shifts is the transverse focusing of the 
polaritons to a cross section comparable to $\lambda^2$ and sufficiently long
dephasing times. 

The authors acknowledge financial support from the European Union within
the network QUACS and the Marie-Curie Trainings-site at the University
of Kaiserslautern as well as the 
German Science Foundation.


\begin{thebibliography}{99}

\bibitem{EIT} S. E. Harris, Phys. Today {\bf 50}(7), 36 (1997);
M. D. Lukin, Rev. Mod. Phys. {\bf 75}, 457 (2003);

\bibitem{resNLO} H. Schmidt and A. Imamo\u glu, Optics Letters \textbf{21}, 1936
(1996); S. E. Harris and Y. Yamamoto, Phys. Rev. Lett. {\bf 81}, 3611 (1998);
M. D. Lukin, P. R. Hemmer, and M. O. Scully,
Adv. At. Mol. Opt. Phys. {\bf 42}, 347 (2000).

\bibitem{Harris} S.E. Harris and L.V. Hau, Phys. Rev. Lett. \textbf{82}, 4611 (1999); 


\bibitem{FleischhauerPRL-2000} M. Fleischhauer and M.D. Lukin, Phys.\ Rev.\
Lett.\ {\bf 84}, 5094 (2000).

\bibitem{LukinPRL:00} M.D. Lukin and A. Imamo\u glu, Phys. Rev. Lett. \textbf{84},
1419 (2000).

\bibitem{JakschPRL-1999} D. Jaksch et. al., Phys. Rev. Lett. \textbf{81}, 3108
(1998).

\bibitem{Bloch-exp} M. Greiner, O. Mandel, T. Esslinger, T.W.~H\"ansch, and
I. Bloch, Nature \textbf{415}, 39 (2002).

\bibitem{HauNature:99} L.V.~Hau, S.E.~Harris, Z. Dutton, and C.H.~Behroozi,
Nature \textbf{397}, 594 (1999).

\bibitem{HauNature:01} C. Liu, Z. Dutton, C.H. Behroozi and L.V. Hau, Nature {\bf %
409}, 490 (2001).


\bibitem{PhillipsPRL:01} D.F. Phillips,
A. Fleischhauer, A. Mair, R.L. Walsworth and M.D. Lukin, 
Phys.\ Rev.\ Lett.\ {\bf 86}, 783 (2001).




\end{thebibliography}
\end{document}